\documentclass{elsart}

\usepackage{graphicx, subfigure}
\usepackage{psfrag}

\begin{document}

\begin{frontmatter}


\title{NMscatt: a program for calculating inelastic scattering
 from large biomolecular systems
 using classical force-field simulations}


\author[nic]{Franci Merzel\corauthref{cor}},
\ead{franc@cmm.ki.si}
\author[ill,delft]{Fabien Fontaine-Vive} and
\author[ill]{Mark R. Johnson\corauthref{cor}}
\corauth[cor]{Corresponding authors.}
\ead{johnson@ill.fr}

\address[nic]{National Institute of Chemistry, Hajdrihova 19, 1000
  Ljubljana, Slovenia}
\address[ill]{Institute Laue Langevin, BP156, 38042 Grenoble cedex
 9, France}
\address[delft]{Radiation, Reactors and Radionuclides Department,
  Faculty of Applied Sciences, Delft University of Technology,
  Mekelweg 15, 2629 JB Delft, The Netherlands}

\begin{abstract}
Computational tools for normal mode analysis, which are widely used in
physics and materials science problems, are designed here in
a single package called \texttt{NMscatt} (Normal Modes \& scattering)
that allows arbitrarily large systems to be
handled. The package allows inelastic neutron and X-ray scattering
observables to be calculated, allowing comparison with experimental
data produced at large scale facilities. Various simplification
schemes are presented for analysing displacement vectors, which are
otherwise too complicated to understand in very large systems.
\end{abstract}

\end{frontmatter}

%{\bf Program summary}\\
%{\it Title of the progam:}\texttt{NMscatt}\\
%{\it Catalog identifier:}\\
%{\it Program summary URL:}\\
%{\it Program obtainable from:} CPC Program Library, Queen's University
%  of Belfast, N. Ireland\\
%{\it Computer:} x86 PC\\
%{\it Operating system:} GNU/Linux, UNIX\\
%{\it Programming language used:} FORTRAN77\\
%{\it Memory required:} Depends on the system size to be simulated.\\
%{\it No. of bits in a word:} 32 or 64\\
%{\it No. of processors used:} 1\\
%{\it Parallelized?:} No\\
%{\it No. of lines in distributed program, including test data, etc.:}
%2 500\\
%{\it No. of bytes in distributed program, including test data, etc.:}
%74 333\\
%{\it Distribution format:} tar.gz\\
%{\it Typical running time:} About 7 hours per one k-point evaluation
%in sampling all modes dispersion curves for a system containing 3550 
%atoms in the unit cell on AMD Athlon 64 X2 Dual Core Processor 4200+.\\
%{\it Nature of the physical problem:} Normal mode analysis, phonons
%calculation, derivation of incoherent and coherent inelastic
%scattering spectra.\\
%{\it Method of solution:} Full diagonalization (producing 
%eigen-vectors and eigen-values) of dynamical matrix which is
%obtained from potential energy function derivation using finite
%difference method.\\

% main text
\section{Introduction}
At large scale facilities for neutron and X-ray scattering, large
quantities of experimental data are produced. For complex, nanoscale
systems, understanding this data requires computer models. In the case
of inelastic scattering, molecular dynamics (MD) simulations
\cite{leach} are
widely used to equilibrate structures and explore dynamics as a
function of temperature and other experimental parameters. However MD
only gives a partial description of vibrational modes through the
partial density of states and when knowledge about specific
vibrational modes is required, normal mode analysis (NMA) \cite{dusa} 
has to be
performed. For physics and materials science problems, NMA gives a
description of the lattice dynamics via the dispersion (k-vector
dependence) of the mode frequencies \cite{frank}. For small systems 
($<$ 200 atoms in
the simulation box) very accurate results can be obtained using
density functional theory (DFT) methods to determine interatomic force
constants \cite{parlinski1, mark1}. By combining DFT and software that 
constructs and
diagonalises the dynamical matrix and calculates the experimental
observables, experimentalists now have sophisticated tools to analyse
their data. The PHONON \cite{phonon} package is one of the best examples. 

Phonon codes are traditionally limited to small systems for a number
of reasons. For example, for small unit cells, the reciprocal lattice
is big and stronger effects of dispersion are expected. If DFT methods
are used to determine force constants then these methods are
themselves restricted to a few hundred atoms. However the development
of nanoscale structures of (partially) crystalline materials
stimulates a need for phonon codes to be extended to much larger
systems. Parameterised force fields \cite{ff} can be used to determine the
inter-atomic force constants and, as will be seen in the example
presented here, strong dispersion effects are observed inspite of the
small reciprocal lattice. 

In biomolecular systems, the need for NMA has long been recognised and
codes like CHARMM \cite{charmm} allow the gamma point normal modes to
be calculated for moderately big systems. In addition, the neutron 
scattering quantities can be directly calculated from the
simulations using the time-correlation function formalism
\cite{vhove}, as implemented in the nMOLDYN program \cite{nmoldyn}.
A combination of neuton scattering experiments and atomic detail computer
simulations has proven to be a powerful technique for studying
internal molecular vibrations \cite{jeremy1,jeremy2,jeremy3}. In this approach 
one can validate the applied numerical models, i.e. force field 
parametrizations, depending on the agreement between the experimental
and calculated spectra.

In this paper we present a software package that extends the
functionality of codes like PHONON \cite{phonon} and Climax 
\cite{climax} to arbitrarily large systems and
extends the gamma point only analysis already available for larger
systems to include k-vector dependence. The software reads a Hessian
matrix of force constants, constructs and diagonalises the dynamical
matrix for any k-vector and calculates neutron and X-ray scattering
observables. The computational bottleneck remains the diagonalisation
of correspondingly large dynamical matrices and we comment on
approximations that have to be used when the Brillouin zone cannot be
sampled at a large number of points. In large systems, atomistic
detail in the displacement vectors can be difficult to interpret due
to the large number of degrees of freedom and we present two methods
for simplifying this information. The first entails summing
displacement vectors over atoms in user-defined beads, while the
second involves a reduction of the degrees of freedom in the dynamical
matrix by summing over force constants, which has the advantage of
reducing the number of modes to be examined. 

\section{Theoretical background}

The standard approach, also called a direct method \cite{parlinsk2},
to the lattice vibration problem of
crystals is based on the explicit knowledge of the interaction 
between all atom-pairs in the system. Subsequently, one deduces 
the corresponding force constants, and constructs and diagonalizes the
dynamical matrix for any k-vector in order to obtain the frequencies
of the normal modes. A reasonable atomic detail description of interactions 
within large biomolecular systems are provided using empirical 
force fields. 

In the following we will briefly summarize the aspects
of the classical theory of lattice vibrations \cite{latt} and proceed to the
description of the explicit phonon calculations.

The individual atomic positions in the crystal can be assigned as
\begin{equation}
\vec R_{n\mu}(t) = \vec R_n+\vec r_\mu+\vec u_{n\mu}(t),
\end{equation}
where $\vec R_n$ is unit cell lattice vector and $\vec u_{n\mu}(t)$ is
displacement of atom $\mu$ from its equilibrium positions $\vec
r_\mu$. Within the harmonic approximation we concentrate on expansion 
of the small differences of potential energy $V$ due to the
small changes in atom positions:
\begin{equation}
V({\bf u})\approx V_0+\sum_{n\mu\alpha}\frac{\partial V}{\partial
  u_{n\mu\alpha}}u_{n\mu\alpha}+\frac{1}{2}\sum_{n\mu\alpha,m\nu\beta}
u_{n\mu\alpha}{\bf D}_{n\mu\alpha,m\nu\beta}u_{m\nu\beta}+\dots,
\end{equation}
where the second derivative defines the force constant between
the atoms $\mu$ and $\nu$:
\begin{equation}
{\bf D}_{n\mu\alpha,m\nu\beta}=\frac{\partial^2 V}{\partial
  u_{n\mu\alpha}\partial u_{m\nu\beta}},\qquad(\alpha,\beta=x,y,z)
\label{le1}
\end{equation}
As each unit cell is identical to every other unit cell in the
crystal, the displacement pattern of a normal mode has to be identical
to that in any other cell to within a phase difference $\vec
k(\vec R_n-\vec R_m)$. The representation of the atom displacement 
is chosen to be a plain wave ansatz of the form: 
\begin{equation}
\vec u_{n\mu}(\vec k,t) = \frac{u_0}{\sqrt{M_\mu}}\vec e_{\mu\vec
  k}\exp(i[\vec k\vec R_{n}-\omega_{\vec k}t]),
\label{le2}
\end{equation}
where $\vec e_{\mu\vec k}$ is the polarization vector and $M_\mu$ is
the mass of the atom $\mu$. We omit writing
Cartesian component subscripts. Solving the equation of motion with ansatz 
(\ref{le2}) is equivalent to the eigen-value problem
\begin{equation}
\omega_{\vec kj}^2\vec e_{\mu\vec kj}=\sum_\nu{\cal D}_{\mu \nu}(\vec k)\vec
e_{\nu\vec kj},
\label{le3}
\end{equation}
where 
\begin{equation}
{\cal D}_{\mu \nu}(\vec k)=\sum_m\frac{1}{\sqrt{M_\mu M_\nu}}{\bf
  D}_{n\mu,m\nu}\exp[i\vec k(\vec R_m-\vec R_n)]
\label{le4}
\end{equation}
is the so called dynamical matrix.

The form of the dynamical matrix (\ref{le4}) requires 
the atom pairs for which one atom belongs to a different unit cell,
{\it i.e.} $(m\ne n)$, to be identified. These terms contribute to the so called
Bloch-factor $\exp[i\vec k(\vec R_m-\vec R_n)]$ and make the dynamical 
matrix complex. But in case of applying periodic boundary conditions
(PBC) as implemented in computer simulation programs, the potential
energy of a crystal is given as an explicit function of only the atom
positions in the primary unit cell. As a consequence, we obtain the 
second derivative matrix $\bf H$ in which the contributions from the
inter-cell atomic pairs are mapped and added to the corresponding image atom 
pairs in the primary unit cell:
\begin{equation}
\left[\sum_m{\bf D}_{n\mu,m\nu}\right]\to{\bf H}_{\mu,\nu}=
\frac{\partial^2 V^{PBC}}{\partial
  u_{\mu}\partial u_{\nu}}
\label{le5}
\end{equation}
One can directly obtain ${\bf D}_{n\mu\alpha,m\nu\beta}$ by 
increasing the size of the unit cell by one or more layers of
periodically arranged image cells and calculate force constants
in the extended supercell. However, this approach is unfavourable when
dealing with very large systems. 

A similar approach is to decompose the potential energy $V^{PBC}$ in
equation (\ref{le5}) into individual contributions from the image
cells, $V^{PBC}=V_0+\sum_mV_m$ and evaluate the second derivative
matrix for each term upon the same minimized structure.

The situation is simpler if the interaction is truncated at some
cutoff distance $r_{cut}$ so that the ``minimum image
convention'' (MIC) is obeyed.  The MIC states that each atom interacts 
at most with one image of every other atom in the system (which is
repeated to fully enclose the primary unit cell with the periodic
boundary conditions). This has the effect of limiting the interaction
cutoff, for example, to no more than half the length of the minimum side when
simulating the orthorhombic cell, $r_{cut} < min\{a/2,b/2,c/2\}$. It 
should be noted
that the size of nanoscale crystals usually far exceeds the 
spatial range of forces between atoms ($\sim$ $<$12 \AA) allowing physically 
reasonable cutoff radii to be introduced.  

\begin{figure}[ht]
\begin{center}
\includegraphics[width=0.5\textwidth]{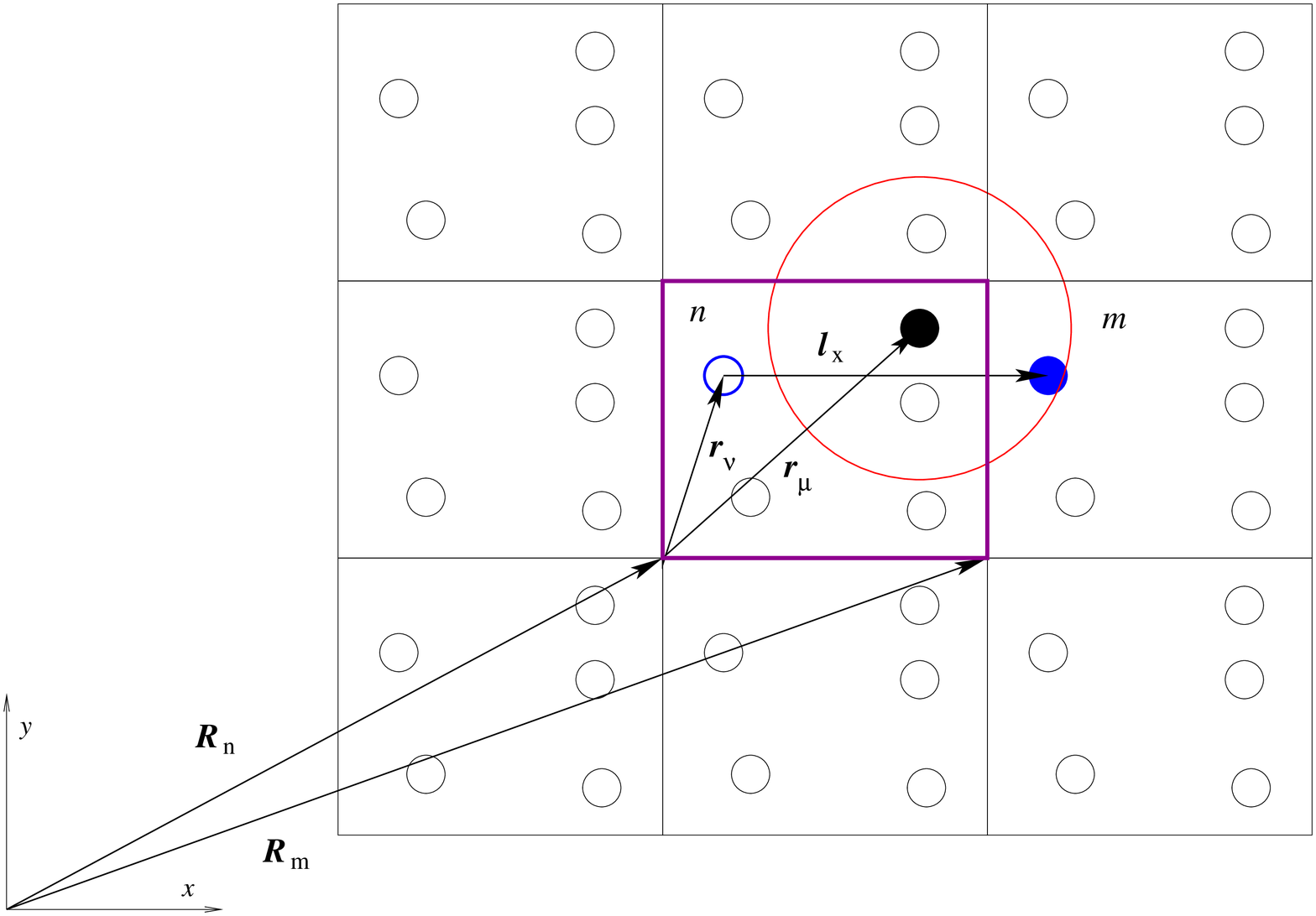}
\caption{A 2D periodic system representing the effect of the minimum
image convention while constructing dynamical matrix.
If a distance between atom $\nu$ (hollow blue circle) and 
$\mu$ (full black circle) is larger than the cut-off distance (red 
circle around the atom $\mu$), and if the force constant 
${\bf H}_{\mu\nu}$ turns out to be nonzero, it means that the 
value of ${\bf H}_{\mu\nu}$ has to be related to the atom $\nu$ in the
image cell $m$ (full blue circle), with the position vector 
$\vec r_\nu+\vec l_x$ ($\vec l_x=\vec R_m-\vec R_n$), instead of to
the atom $\vec r_\nu$ in the primary cell $n$. The corresponding lattice
translation vector $\vec l_x$ is thus determined for the element ${\bf
H}_{\mu\nu}$.} 
\label{fg1}
\end{center}
\end{figure}

According to the MIC we can conclude that there is always only one
translation $\vec t$ per atom pair $(\nu,\mu)$ giving rise to the 
minimum distance $|(\vec r_\nu+\vec t)- \vec r_\mu|$:
\begin{eqnarray}
\exists s_\alpha &=& \{-1,0,1\}_{\alpha=1,2,3}: 
\vec t=s_1\vec l_1 + s_2\vec l_2 + s_3\vec l_3;\\
&& {\textrm{so
    that:}\quad |(\vec r_\nu+\vec t)- \vec r_\mu|=min,}\nonumber 
\end{eqnarray}
where $\vec l_\alpha$ are lattice translation vectors, c.f. Fig. \ref{fg1}.
In the case of a nonvanishing ${\bf  H}_{\mu,\nu}$ we get the
following expression for a dynamical matrix element 

\begin{equation}
{\cal D}_{\mu \nu}(\vec k)=\frac{1}{\sqrt{M_\mu M_\nu}}\left\{
\begin{array}{ll}
{\bf H}_{\mu,\nu}, & {\textrm{if : }} |\vec r_\mu-\vec
r_\nu|<r_{cut}\\
{\bf H}_{\mu,\nu}\exp(i\vec k\vec t), & {\textrm{if : }} |\vec r_\mu-\vec
r_\nu|>r_{cut}\\
\end{array}\right.
\end{equation}

According to equation (\ref{le3}) the diagonalization of 
matrix ${\cal D}(\vec k)$ yields the phonon
frequencies $\omega_{\vec kj}$ and corresponding polarization
vectors $\vec e_{\mu\vec kj}$ for a given phonon wave vector $\vec k$.
A complete solution leads to the phonon dispersion relations. The
subscript $j$ denotes a branch in the phonon dispersion. In a
crystal of $N$ atoms, there are $3N$ branches.
 
\vskip 0.2cm
{\bf Inelastic scattering}
\vskip 0.2cm

The dynamic structure factor $S(\vec q,\omega)$ contains information
about the structure and dynamics of the sample. It can be split into
a coherent part arising from the cross-correlations of atomic motions
and an incoherent part describing self-correlations of single atom
motions. According to the standard theory \cite{bee, lovesey},
which is based on the harmonic approximation, we obtain the following 
expressions for the coherent and incoherent dynamical structure factors:
\begin{eqnarray}
S(\vec q,\omega)_{coh} &=& \sum_{\vec G}\sum_{\vec
  k,j}\frac{\hbar}{2\omega_{\vec kj}}
\left|\sum_\mu\sigma_\mu^{coh}\frac{\vec q \cdot\vec e_{\mu\vec
  kj}}{\sqrt{M_\mu}}\exp(-W_\mu(\vec q)+i\vec q\vec r_\mu)
\right|^2\times\nonumber\\
&\times&
(n(\omega_{\vec kj})+1)\delta(\omega-\omega_{\vec kj})
\delta(\vec q+\vec k-\vec G),
\label{eqn_coh}
\end{eqnarray}
and
\begin{eqnarray}
S(\vec q,\omega)_{inc} &=& \sum_\mu\sigma_\mu^{inc}\sum_{\vec
  k,j}\frac{\hbar}{2M_\mu\omega_{\vec kj}}|\vec q \cdot\vec e_{\mu\vec
  kj}|^2(n(\omega_{\vec kj})+1)\times\nonumber\\
&\times& \exp[-2W_\mu(\vec
  q)]\delta(\omega-\omega_{\vec kj}),
\end{eqnarray}
where $\vec q$ is the scattering vector, $\sigma_\mu$ is the
corresponding atomic scattering
length, $(n(\omega)+1)$ refers to the 
phonon creation process (absorption spectrum) and $n(\omega)$ is the 
mean number of phonons of frequency $\omega$ at temperature $T$ 
according to the Bose-Einstein statistics
\begin{equation}
(n(\omega)+1)=\frac{\exp(\hbar\omega/k_BT)}{\exp(\hbar\omega/k_BT)-1}
\end{equation}
The factor $\exp[-2W_\mu(\vec q)]$ is called the Debye-Waller factor:
\begin{equation}
W_\mu(\vec q)=\frac{\vec{q}\cdot{\bf B}(\mu)\cdot\vec{q}}{2},\qquad
B_{\alpha\beta}(\mu)=\langle u_{\mu\alpha}u_{\mu\beta}\rangle_T,
\end{equation}
where ${\bf B}(\mu)$ is a $3\times 3$ symmetric tensor representing
the thermodynamic mean square displacement of an atom $\mu$, which can
be expressed by the partial atomic phonon density of states 
$g_{\alpha\beta,\mu}(\omega)$:
\begin{equation}
B_{\alpha\beta}(\mu)=\frac{{3N}\hbar}
    {2M_\mu}\int_0^{\infty}\frac{d\omega}{\omega}
    g_{\alpha\beta,\mu}(\omega)
    \coth\left(\frac{\hbar\omega}{2k_BT}\right).
\label{le6}
\end{equation}
The partial atomic density of states is a weighted distribution of
normal modes
\begin{equation}
g_{\alpha\beta,\mu}(\omega)=\frac{1}{3Nn}\sum_{\vec
  k,j}^{n3N}e_{\alpha\mu\vec kj}e^*_{\beta\mu\vec kj}
\delta(\omega-\omega_{\vec kj}),
\end{equation} 
where $n$ is the number of sampling $\vec k$-points in the first
Brillouin zone.

The evaluation of the Debye-Waller $B$ factors
using equation (\ref{le6}) requires extra attention due to the
``zero-phonon'' term resulting from 
the singularity caused by phonon (acoustic) branches where 
$\omega(\vec q,j)=0$. The contributions of the 3 acoustic
modes is treated separately by using the Debye approximation for the
density of states, {\it i.e.} $g_{ac}(\omega)\propto\omega^2$, and
normalization $\int_0^{\omega_{max}}g_{ac}(\omega)d\omega=3$, 
where $\omega_{max}$ is the maximum frequency up to which the 
acoustic dispersion curve is linear
\begin{eqnarray}
B_{\alpha\beta}(\mu) &=& \sum_{j=1}^3\frac{3\hbar}
    {M_\mu\omega_{j\,max}^3}\int_0^{\omega_{j\,max}}
    \omega e_{\alpha\mu\vec kj}e^*_{\beta\mu\vec kj}
    \left(\frac{1}{\exp(\hbar\omega/k_BT)-1}+\frac{1}{2}\right)d\omega
    \nonumber\\
&=& \sum_{j=1}^3\frac{3\hbar e_{\alpha\mu\vec kj}e^*_{\beta\mu\vec kj}}
    {M_\mu\omega_{j\,max}}\left[\left(\frac{k_BT}
    {\hbar\omega_{j\,max}}\right)^2\int_0^{x_{j\,max}}
    \frac{x}{\exp(x)-1}dx+\frac{1}{4}\right].
\end{eqnarray}
A new variable $x_{j\,max}={\hbar\omega_{j\,max}}/{k_BT}$ was
introduced in the last equation.

\section{Analysing the displacement vectors}

For systems containing thousands of atoms (N) the displacement vectors
obtained by diagonalising the dynamical matrix can be difficult to
understand, especially for low frequency modes which involve the
displacement of many (or all) atoms. One simple solution to this
problem is to sum over the displacements of atoms within beads, which
represent logical coarse grains of the system, for example base
molecules in the DNA example below. This treatment allows different
bead definitions to be applied to the calculated displacement
vectors but has the disadvantage of not reducing the number of
displacement vectors from 3N. 

A related approach is to reduce the atomic level Hessian matrix to
lower dimension by mapping the inter-atomic force constants on to
inter-bead force constants \cite{venkat}. For N' beads, the resulting dynamical
matrices have dimension 3N' and therefore result in 3N' displacement
vectors for any k-vector. We note that any reduction in the
dimensionality of the system causes a loss in information, which is
the rotational degrees of freedom of the beads (rigid bodies).

\section{Example}

To verify the implemented formalism we have simulated a
B-form DNA molecule (right-handed, 10 base-pairs per turn, pitch
33.6\AA) using CHARMM \cite{charmm}. The full crystal
environment was generated using periodic boundary conditions for
an orthorhombic unit cell containing one helix of DNA.
The dimension of the unit cell is $a=32.2$\AA, $b=31.8$\AA\  and
$c=33.5$\AA, with $c$ parallel to the helical axis.
The starting configuration was obtained by minimization of the
potential energy of the crystal structure obtained as the
time average over a 1ns MD simulation at 100K. The Hessian matrix of
force constants was generated by displacing each atom in turn from
equilibrium and calculating the forces induced on all other
atoms. Diagonalisation of the resulting dynamical matrices was
performed using the the routine {\texttt{zcheev}} from the LAPACK
library \cite{LAPACK}. 
 
\begin{figure}
\psfrag{S(q,w) [arb. units]}{\small $S(\vec q,\omega)$ [arb. units]}
\psfrag{energy transfer [cm-1]}{\small energy transfer $\omega [cm^{-1}]$}
\begin{center}
\includegraphics[width=0.8\textwidth]{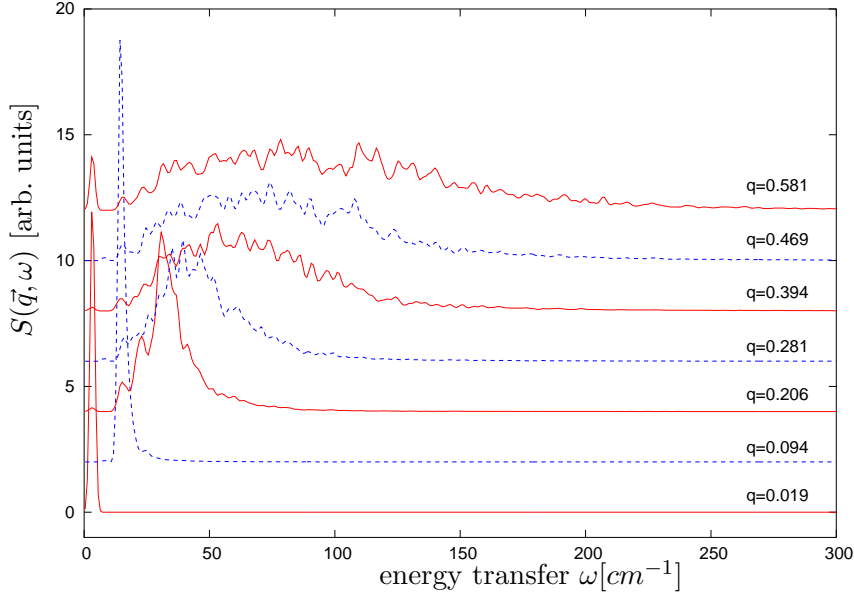}
\caption{Coherent dynamical structure factor of B-DNA. 
Scattering vector $\vec  q$ is varied along the helical axis (0,0,1). 
Shown are $S(\vec q,\omega),\ |\vec q|=0.019\dots0.581$\AA$^{-1}$.}
\label{fg2}
\end{center}
\end{figure}

\begin{figure}
\psfrag{w [cm-1]}{\small $\omega$ [cm$^{-1}$]}
\psfrag{q [A-1]}{\small $q$[\AA$^{-1}$]}
\begin{center}
\includegraphics[width=0.8\textwidth]{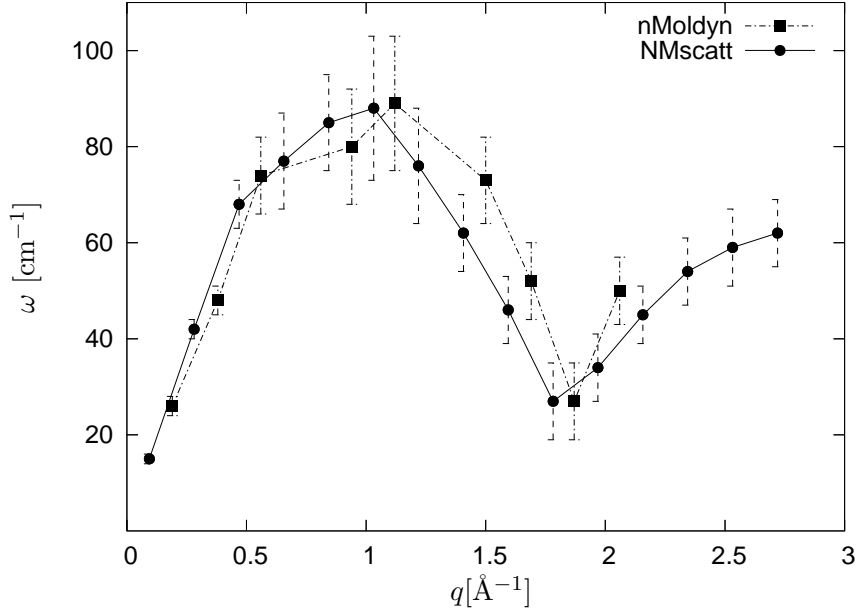}
\caption{Dispersion curves obtained by Gaussian fitting of the
  calculated Brillouin peak for inelastic X-ray scattering using 
\texttt{NMscatt} (solid curve + circles), and using nMoldyn (dashed
  curve + squares).}
\label{fg3}
\end{center}
\end{figure}

The typical coherent spectrum of DNA, Figure \ref{fg2}, obtained from
equation \ref{eqn_coh}, shows a well-defined 
Brillouin peak at small $\omega$ which moves along the frequency 
dimension upon varying momentum transfer $\vec q$. Fitting the
spectral profile with a Gaussian as a function of wave-vector gives
the dispersion curve shown in Figure 
\ref{fg3}, which compares well with the recent experimental 
results \cite{dna_jcp}.
Figure \ref{fg3} also shows the result of the equivalent analysis of a
300K MD simulation on the same model of B-DNA using nMoldyn. 

\begin{figure}
\begin{center}
\subfigure[B-DNA: acoustic mode]{\includegraphics
[width=0.40\textwidth]{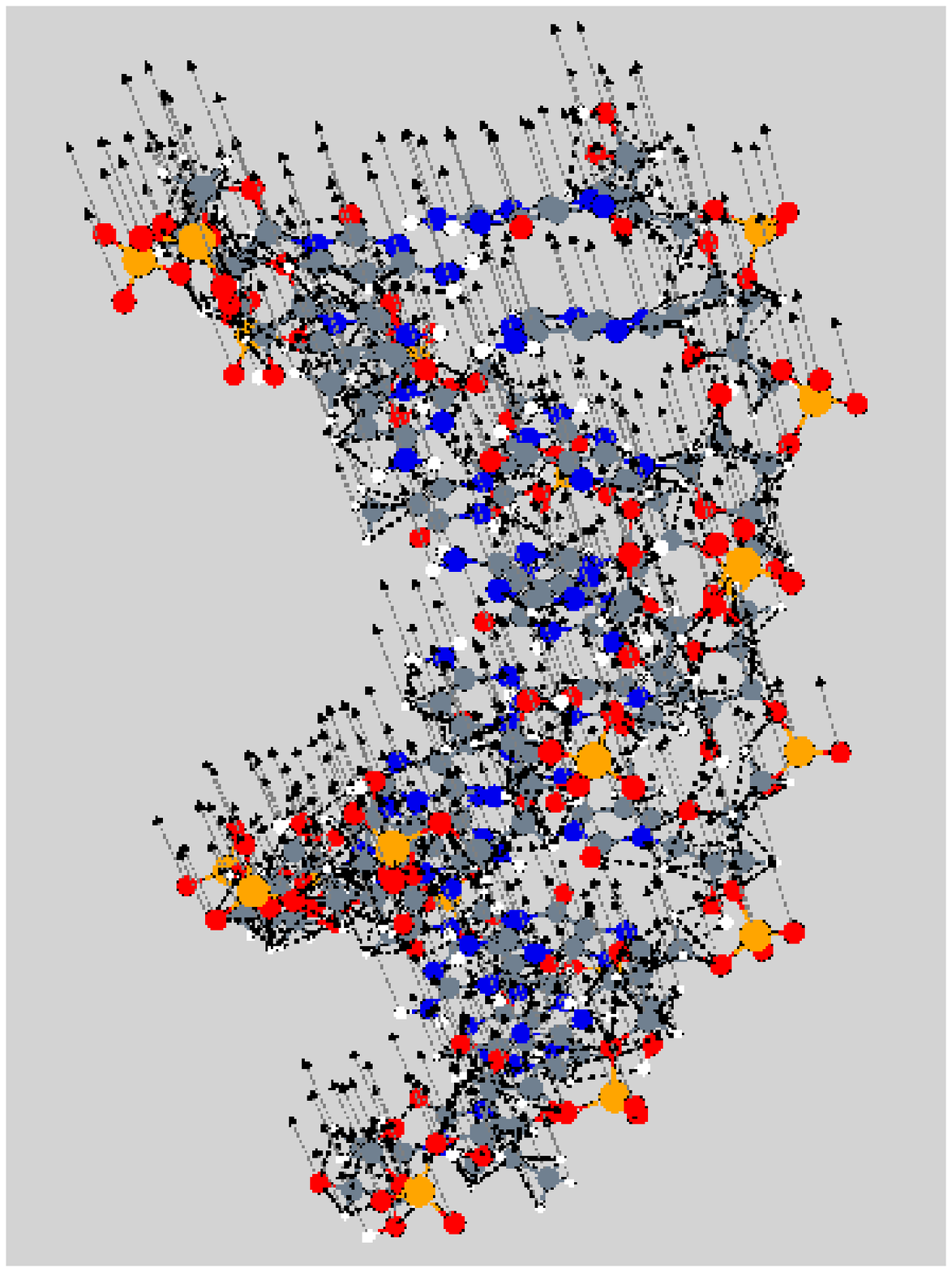}}
\qquad
\subfigure[B-DNA: beads]{\includegraphics
[width=0.454\textwidth]{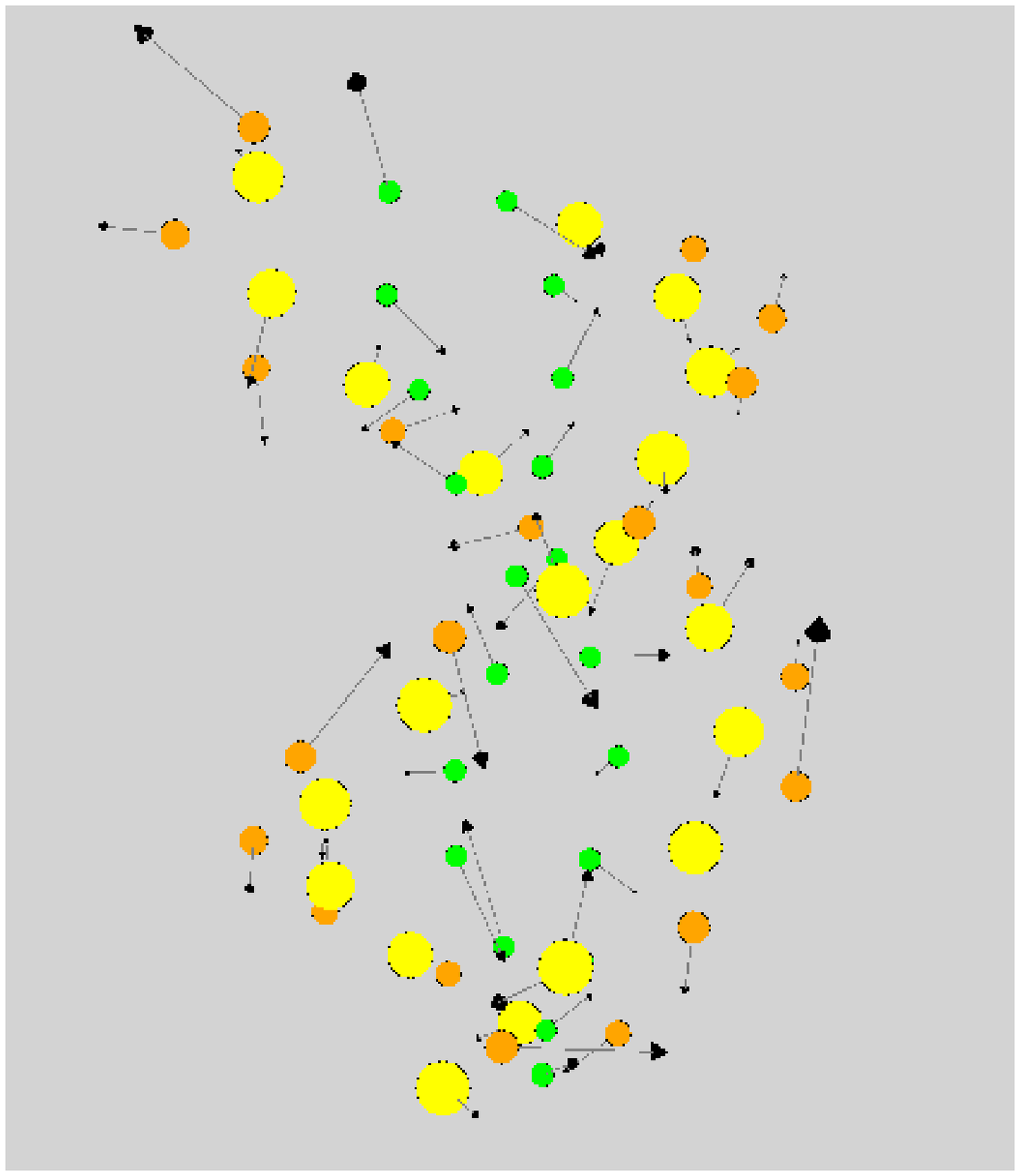}}
\caption{Atomic {\it a)} and {\it b)} beads 
(base molecules, sugar molecules and
  phosphate groups) displacements of the selected mode.} 
\label{fg5}
\end{center}
\end{figure}

In order to gain insight into the nature of low frequency dynamics we
can analyse the displacement vectors at an atomic level (see Figure
\ref{fg5}(a) for an acoustic mode). 
By summing over the displacement
 vectors in terms of beads, where base
molecules, sugar molecules and phosphate groups are treated as single
units, a simplified picture of the normal modes is obtained. Figure
\ref{fg5}(b) shows a high frequency mode which has a pronounced
contribution from the phosphate groups. 

\section{Conclusion}
The new, user-friendly computational package
\texttt{NMscatt} presented here enable an efficient
atomic detail analysis of different types of inelastic scattering
applied to arbitrarily large nanoscale systems. The ability to
perform molecular dynamics and phonon calculations on large
nano and bio-molecular materials means that one can efficiently pursue
the investigation of some poorly understood structural and dynamical 
features of these systems.
 
 \section{Acknowledgements.} 
The authors are grateful to
Dr. Stephane Rols for helpful discussions at the start of this project. 
MJ acknowledges a long-standing collaboration with Prof. Krzysztof
Parlinski. FM acknowledges a financial support from ILL during his
stay in Grenoble and support from the Ministry of Higher Education, 
Science and Technology of Republic of Slovenia under Grant nos. 
P1-0002, J1-6331 and J1-5115.

\vskip 1cm
\appendix
\section{Program package and data structure}

\begin{figure}[ht]
\begin{center}
\includegraphics[width=0.7\textwidth]{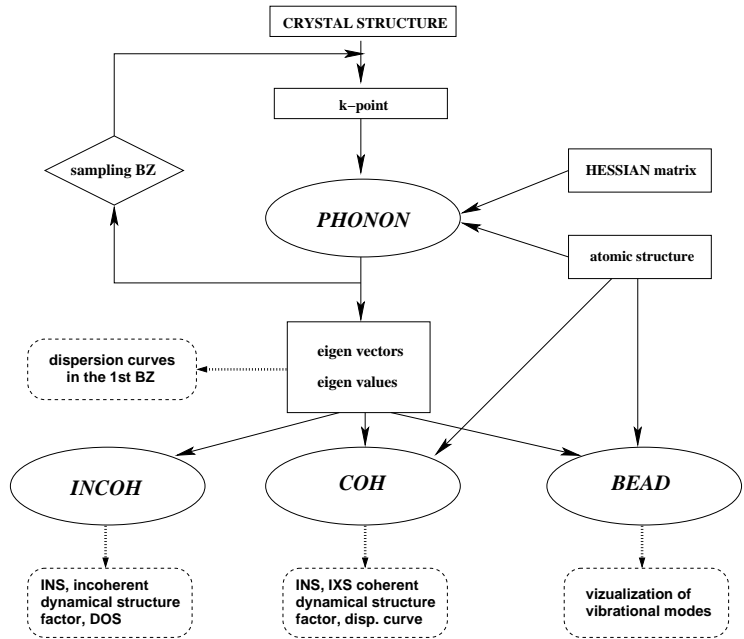}
\caption{Flow-chart of the \texttt{NMscatt} program package. }
\label{fg6}
\end{center}
\end{figure}

There are four main modules in the \texttt{NMscatt} program package
{\it phonon, coh, incoh} and {\it bead}, and the overall
\texttt{NMscatt} structure
is given in Figure \ref{fg6}. Below are described the corresponding
modules.
\begin{itemize}
\item{\it phonon}: Providing the full Hessian matrix for a given 
energy-minimized atomic structure within the specified
crystallographic unit cell this module constructs dynamical matrix and
calculates its eigenvalues and eigenvectors at given wave vector
$\vec k$. At input this module requires to specify the Bravais lattice
vectors that were previously used in the molecular mechanics/dynamics
simulation 
package to satisfy the periodic boundary conditions while generating
the minimized structure. The cut-off radius must be given at which 
the long range interactions are truncated while calculating the Hessian 
matrix in the simulation. At output two separate files 
\texttt{eig\_val\_}$n$ and
\texttt{eig\_vec\_}$n$ are written containing
eigenvalues and eigenvectors, respectively, where $n$ assigns 
a consecutive number of the sampling $k$-point in the first Brillouin
zone. These files serve as an imput for other modules of
\texttt{NMscatt}. The choice $n=0$ is assumed to be reserved for the
$\Gamma$-point, $\vec k=(0,0,0)$. The lowest $n$ should sample
the vicinity of the $\Gamma$-point. The $k$-points are to be specified
in the fractional coordinates with respect to the reciprocal lattice
vectors. The elements of the upper triangle of the hessian matrix should be
provided in the binary (default) \texttt{hessian\_uf} or alternatively in the
ASCII file \texttt{hessian\_f} and the atomic coordinates in the file 
\texttt{coord} writen in the CHARMM coordinate format. 
 
\item{\it incoh}: This module allows to calculate atomic
Debye-Waller factors representing the mean square displacements
and dynamical structure factor of incoherent one-phonon
neutron scattering on monocrystals and from orientationally 
averaged powder. In the latter case we need to specify an
absolute value of momentum transfer $q$, range of the $k$-point
index $n$: $0-n_{max}$ for picking up the corresponding $k$-point
eigenvalues- and eigenvectors-files generated by {\it phonon}, 
number of random orientations
of $q$ vector to provide spherical averaging and absolute temperature.
Also, one has to define the $k$-point index $n_a$, for which the
$k$-point range $0-n_a$ corresponds to the linear regime of the
acoustic-mode-dispersion curves. This is needed for the proper 
derivation of Debye-Waller factors using Debye approximation.
As a result, the function $S(q,\omega)$ is given in the file 
\texttt{s\_q\_w}. Optionally, one can also obtain density of states
(DOS) in this module.
  
\item{\it coh}: This module calculates dynamical structure factor of 
coherent one-phonon neutron or X-ray scattering on monocrystals. 
In addition to the input data required by {\it incoh} (except for
the spherical averaging), we need to define for module {\it coh} also
the type of scattering neutron/X-ray, the number of higher Brillouin
zones included for sampling momentum transfer vector, and the range
of the $k$-point index $n_1-n_2$ evaluated by module {\it phonon} 
and assigning the $k$-points in the first Brillouin zone, which lie 
along the selected direction of the momentum transfer vector $\vec q$.  
As a result, the function $S(q,\omega)$ is given in the file 
\texttt{s\_qw\_coh}. 

\item{\it bead}: This module is used to enable visualization of selected
vibrational modes obtained by running {\it phonon} for the $\Gamma$-point,
such that atomic displacement vectors are projected on to the beads,
which are defined as the center of mass of larger atomic groups
of the system, for example residues. The output files are readable
by the program \texttt{xmakemol}\cite{xmkml} which enables direct
visualisation of the mode.

\end{itemize}

\pagebreak
\noindent
{\bf Compiling}

To compile the program package Makefile is provided. It is important
to note that 64-bits processors are prerequisite for applying the
\texttt{NMscatt} to analysis of the large systems (containing more
than 2000 atoms). One needs to install the LAPACK library on the 
computer beforehand,  prior to \texttt{NMscatt}. The fortran compiling 
switches \texttt{g77 -mcmodel=medium -funroll-all-loops -fno-f2c -O3} 
are recomended when installed on 64-bits processors running Linux.

\noindent
{\bf Benchmark results}

Bencmark results were obtained on AMD Athlon 64 X2 Dual Core Processor
2.2GHz running Linux for B-form DNA simulated with the CHARMM program.

%\noindent
%{\bf Sample input}

% \appendix

% \section{}
% \label{}

%\begin{thebibliography}{00}


% notes:

% \bibitem{label2}
% If there is a note, it should come last:
% \bibitem{label3} \note
% \end{subbibitems}

%\bibitem{}

%\end{thebibliography}

\end{document}